\documentclass[utf8,a4paper]{scrartcl}

\usepackage{url,hyperref,lineno,microtype,subcaption}
\usepackage[onehalfspacing]{setspace}

\usepackage{graphicx}%
\usepackage{multirow}%
\usepackage{amsmath,amssymb,amsfonts}%
\usepackage{amsthm}%
\usepackage{mathrsfs}%
\usepackage[title]{appendix}%
\usepackage{xcolor}%
\usepackage{textcomp}%
\usepackage{manyfoot}%
\usepackage{booktabs}%

\usepackage[affil-it]{authblk}

\usepackage[T1]{fontenc}

\usepackage{siunitx}
\usepackage{isotope}
\usepackage{graphicx,tabularx,booktabs,colortbl}
\usepackage{pgfplotstable}
\usepackage{subcaption}
\usepackage{hyperref}
\usepackage[citestyle=nature]{biblatex}

\title{First Positronium Lifetime Imaging with Scandium-44 on a Long Axial Field-of-view PET/CT} 

\author[1,2,*]{Lorenzo Mercolli} %\email{lorenzo.mercolli@insel.ch}
\author[3]{William M. Steinberger}%\email{william.steinberger@siemens-healthineers.com}
\author[4,5]{Pascal V.\ Grundler}
\author[4,5]{Anzhelika Moiseeva}
\author[6]{Saverio Braccini}
\author[3]{Maurizio Conti}%\email{maurizioconti@siemens-healthineers.com}
\author[7,8]{Paweł Moskal}%\email{p.moskal@uj.edu.pl}
\author[1,2]{Narendra Rathod}%\email{narendra.rathod@unibe.ch}
\author[1]{Axel Rominger}%\email{axel.rominger@insel.ch}
\author[1,2,9]{Hasan Sari}%\email{hasan.sari@unibe.ch}
\author[5,10]{Roger Schibli}
\author[1,2]{Robert Seifert}
\author[1,2]{Kuangyu Shi}
\author[7,8]{Ewa Ł. Stępień}%\email{e.stepien@uj.edu.pl}
\author[4,5]{Nicholas P.\ van der Meulen}

\affil[1]{Department of Nuclear Medicine, Inselspital, Bern University Hostpital, University of Bern, Bern, Switzerland}
\affil[2]{ARTORG Center for Biomedical Engineering Research, University of Bern, Bern, Switzerland}
\affil[3]{Siemens Medical Solutions USA, Inc., Knoxville TN, USA}
\affil[4]{Laboratory of Radiochemistry, PSI Center for Nuclear Engineering and Sciences, 5232 Villigen-PSI, Switzerland}
\affil[5]{Center for Radiopharmaceutical Sciences, PSI Center for Life Sciences, 5232 Villigen-PSI, Switzerland}
\affil[6]{Albert Einstein Center for Fundamental Physics (AEC), Laboratory for High Energy Physics (LHEP), University of Bern, Bern, Switzerland}
\affil[7]{Faculty of Physics, Astronomy and Applied Computer Science, Jagiellonian University, Krakow, Poland}
\affil[8]{Centre for Theranostics, Jagiellonian University, Krakow, Poland}
\affil[9]{Siemens Healthineers International AG, Z{\"u}rich, Switzerland}
\affil[10]{Department of Chemistry and Applied Biosciences, ETH Zurich, Zurich, Switzerland }
\affil[*]{\url{mailto:lorenzo.mercolli@insel.ch}}

\newcommand{\ps}{\mbox{ps}}
\newcommand{\ns}{\mbox{ns}}
\newcommand{\kev}{\mbox{keV}}

\newcommand{\mbq}{\mbox{MBq}}
\newcommand{\kbqml}{\mbox{kBq}/\mbox{ml}}

\newcommand{\ga}{\isotope[68]{Ga}}
\newcommand{\iod}{\isotope[124]{I}}
\newcommand{\rb}{\isotope[82]{Rb}}

\newcommand{\scan}{\isotope[44]{Sc}}

\newcommand{\psma}{[\isotope[68]{Ga}]Ga-PSMA-617}
\newcommand{\dota}{[\isotope[68]{Ga}]Ga-DOTA-TOC}
\newcommand{\tops}{\tau_{3}}
\newcommand{\tge}{$3\gamma\mbox{E}$}
\newcommand{\imgbin}{$4\times 4 \times 4 \, \mbox{mm}^3$}

\begin{document}

\maketitle

\newpage
\begin{abstract}
{\centering {\sectfont Abstract}}

%%% Leave the Abstract empty if your article does not require one, please see the Summary Table for full details.
\textbf{Purpose: } \textsuperscript{44}Sc has been successfully produced, synthesized, labeled and first-in-human studies were conducted some years ago. The decay properties of \textsuperscript{44}Sc, together with being close to a clinical implementation, make it an ideal candidate for in vivo positronium lifetime measurements. In this study, we investigate the count statistics for ortho-positronium (oPs) measurements with \textsuperscript{44}Sc.

\textbf{Method: } A NEMA image quality phantom was filled with 41.7 MBq of \textsuperscript{44}Sc dissolved in water and scanned on a commercial long-axial field-of-view PET/CT. Three-photon events were identified using a prototype feature of the scanner and dedicated software. The lifetime of oPs was determined in the phantom spheres and in  4x4x4 mm\textsuperscript{3} voxels. 

\textbf{Results:} All measured oPs lifetimes are compatible, within the uncertainties, with the literature values for water. The oPs lifetime is 2.65±0.50, 1.39±0.20 and 1.76±0.18 ns in the three smallest spheres of the phantom and 1.79±0.57 ns for a single voxel in the central region of the largest sphere. The relative standard deviation in the background regions of the time difference distributions, i.e., for time differences smaller than -2.7 ns, is above 20\% - even for voxels inside the phantom spheres.  

\textbf{Conclusions:} Despite the favorable physical properties of \textsuperscript{44}Sc, the count statistics of three-photon events remains a challenge. The high prompt-photon energy causes a significant amount of random three-photon coincidences with the given methodology and, therefore, increases the statistical uncertainties on the measured oPs lifetime. 

\vspace{2em}
{\small\textbf{Keywords:} Scandium-44, long axial field-of-view PET/CT, positronium, positronium lifetime imaging} 
\end{abstract}

\newpage 
%%%%%%%%%%%%%%%%%%%%%%%%%%%%%%%%%%%%%%%%%%%%%%%%%%%%%%%%%%%%%%%%%%%%%%%%%%%%%%%
\section{Introduction}

Investigating the lifetime of ortho-positronium (oPs), the spin-1 state of an electron-positron bound system, has offered valuable insights into the structural properties of matter for decades (see e.g.\ Refs.\cite{cizek_CharacterizationLatticeDefects_2018,gidley_PositronAnnihilationMethod_2006,jean_PerspectivePositronAnnihilation_2013,kobayashi_PositroniumChemistryPorous_2007,schmidt_PositronAnnihilationSpectroscopy_2008,schultz_InteractionPositronBeams_1988,suvegh_PositronAnnihilationSpectroscopies_2011,tuomisto_DefectIdentificationSemiconductors_2013}). More recently, the medical community has shown interest in measuring oPs lifetimes in human tissue \cite{moskal2019b,bass2023,hourlier2024,moskal2025}. So-called \emph{oPs lifetime imaging}, i.e.\ constructing a three-dimensional image of the human body with the oPs lifetime as voxel value~\cite{moskal2019ieee}, has the potential to provide diagnostic information about the tissue microenvironment, in particular oxygenation levels, that is currently unavailable in clinical routine (see e.g.  Refs.~\cite{Axpe2014,Moskal2019,moskal2019ieee,shibuya2020,Stepanov2020,Moskal2021,moskal2022,qi2022,moskal2023,chen2023,takyu2024b}). Recently, the first in vivo oPs lifetime images were determined with the dedicated multi-photon J-PET scanner prototype~\cite{moskal2024brain}, and notably also the first in vivo oPs lifetime measurements with a commercial PET/CT system were demonstrated~\cite{mercolli2024}. The methods for positronium lifetime image reconstruction are also being intensively developed~\cite{qi2022,chen2023,shopa2023,Huang2024b,chen2024,Berens2024,Huang2025,HHHuang2025}.

The oPs lifetime can be measured by determining the time difference between a prompt gamma photon, emitted during the nuclear decay with the positron, and the two photons with $511 \, \kev$ energy from the positron annihilation. The prompt photon serves as the start time, while the detection of the annihilation photons sets the stop time. The two annihilation photons are also used to determine the place of annihilation~\cite{moskal2020a}. Histograming all measured time differences gives a Positron Annihilation Lifetime (PAL) spectrum  that contains several components, including the oPs lifetime. The oPs lifetime is of particular interest, as it depends on the molecular structure of the surrounding matter~\cite{moskal2019b,bass2023}.

oPs lifetime measurements require a positron-emitting radionuclide with prompt-photon emission, together with the possibility of detecting and localizing three-photon events\footnote{In this study, we do not consider three-photon decays of oPs.} (\tge). The detection of \tge\ poses significant challenges, particularly in a clinical environment. Positron emission tomography (PET) systems are designed to detect photon pairs with $511 \, \kev$ energy. The detection of single-photon events with different energies is not part of the core design of clinical PET/CT scanners. Nonetheless, Ref.~\cite{steinberger2024} presented the first use of a clinical PET/CT scanner for oPs lifetime measurements by extending the detection and processing capabilities to \tge. The number of detected \tge\ is critical to achieve an accurate oPs lifetime measurement. The increased sensitivity of long-axial field-of-view (LAFOV) PET/CT systems (see e.g.\ Refs.~\cite{Alberts2021,Prenosil2022,Spencer2021,moskal2020b}) proved to be a key factor for oPs lifetime measurement on a commercial PET/CT system. 

Radionuclides with prompt-photon emission are readily available in clinics, of which \ga\ labeled with \psma\ and \dota\ is by far the most widely adapted. \rb\ and to some extent \iod\ are also used in clinical routine, which is why Refs.~\cite{moskal2024brain,mercolli2024} relied on \ga\ and \rb\ for in vivo measurements. The prompt photon branching ratio (BR) is, of course, a key physical parameter to maximize the count statistics of \tge. As shown in Tab.~\ref{t:isotopes}, \ga\ and \rb\ have only a limited prompt-photon BR. If the positron emission fraction is taken into account, also the seemingly high prompt-photon BR of \iod\ drops significantly. \scan, on the other hand, has a very high prompt-photon BR in conjunction with a high positron fraction, which makes it a prime candidate for oPs lifetime imaging~\cite{moskal2020b,das2023}. There is legitimate hope that \scan\ can overcome the challenge of detecting enough \tge\ for a reliable determination of the useful lifetime of oPs~\cite{moskal2020b}. In this respect, \isotope[43]{Sc} appears to be in the ballpark of \ga, \rb, and \iod\ for oPs lifetime imaging. 

\begin{table}[htb]
    \centering
    \caption{Comparison of \scan's decay properties with other clinically viable radionuclides (retrieved from \url{https://www-nds.iaea.org/}). %Refs.~\citep{nds43,nds44,nds68,nds82,nds124}.
    BR$_{\gamma}/\beta^+$ is the prompt photon BR per positron, i.e.\ without electron capture decays.}
    \label{t:isotopes}
    \resizebox{\textwidth}{!}{
    \begin{tabular}{lcccccccc}
        \toprule
        Nuclide & Decay const. $[\mbox{s}^{-1}]$ & $\langle E_{\beta^+} \rangle \; [\kev]$ & BR $_{\beta^+}$ $[\%]$ & $E_\gamma \; [\kev]$ & BR$_\gamma$ $[\%]$ & BR$_{\gamma}/\beta^+$ $[\%]$ \\
        \midrule
        \rowcolor{lightgray} & & $344.5 \pm 0.8$ & $17.2 \pm 0.5$ &  & & \\
        \rowcolor{lightgray} \multirow{-2}{*}{\isotope[43]{Sc}} &  \multirow{-2}{*}{$(4.948 \pm 0.015)\cdot 10^{-5}$} & $508.1 \pm 0.9$ & $70.9 \pm 0.6$ & \multirow{-2}{*}{$372.9 \pm 0.3$} & \multirow{-2}{*}{$22.5 \pm 0.7 $ }  & \multirow{-2}{*}{$17.2\pm0.8$}\\
        \scan & $ (4.764 \pm 0.003)\cdot 10^{-5}$ & $630.2 \pm 0.8$ & $94.278 \pm 0.011$ &  $1157.022 \pm 0.015$ & $99.887 \pm 0.003 $ & $94.283 \pm 0.023$\\
        \rowcolor{lightgray}  \ga & $ (1.706\pm 0.002) \cdot 10^{-4}$  &  $836.0 \pm 0.6$ & $87.72\pm0.09$  & $1077.37 \pm 0.04$ &  $3.22 \pm 0.03$ & $1.190 \pm 0.017$ \\
        \multirow{2}{*}{\rb} & \multirow{2}{*}{$(9.1868\pm 0.0015)\cdot 10^{-3}$} &  $1169.0\pm 1.4 $ & $13.0\pm0.4$    &   \multirow{2}{*}{$776.511 \pm 0.10$} & \multirow{2}{*}{$15.1 \pm 0.3$ } & \multirow{2}{*}{$13.5 \pm 0.5$} \\
        &  & $1536.0\pm 1.5$  & $81.8 \pm 0.4$ &  & \\
        \rowcolor{lightgray} &  & $687.0\pm0.9$  & $11.7 \pm 1.0$ &  & & \\
        \rowcolor{lightgray} \multirow{-2}{*}{\iod} & \multirow{-2}{*}{$(1.92111 \pm 0.00014)\cdot 10^{-6}$}  & $974.7\pm0.9$ & $10.7\pm 0.9$ & \multirow{-2}{*}{$602.73\pm0.08 $} & \multirow{-2}{*}{$62.9\pm0.7$} & \multirow{-2}{*}{$12.0 \pm 1.1$} \\
        \bottomrule
    \end{tabular}
    }
\end{table}

% Refs.~\cite{steinberger2024,mercolli2025} described the favorable properties of \iod\ in comparison with \ga\ and \rb. %The high prompt-photon BR make \scan\ a prime candidate for oPs lifetime imaging. 
Although \scan\ is not yet available in clinical routine, production routes, purification and labeling as well as first in-human studies have been reported in the literature (see e.g. Refs.~\cite{mueller2014,singh2015,eppard2017,umbricht2017,muller2018,zhang2019,vandermeulen2020,Lima2020,Lima2021,vandereulen2021}). \scan\ can be paired with its therapeutic analog \isotope[47]{Sc} for theranostic applications, enabling seamless transitions between diagnostic imaging and targeted therapy. Adding diagnostic information from oPs lifetime imaging could boost the tailored effectiveness of \isotope[47]{Sc}'s therapeutic application. % In comparison with \iod, the ease of labeling \scan\ with PSMA and DOTA-TOC should be highlighted. 

In this brief report, we investigate the properties of \scan\ for oPs lifetime imaging on a commercial LAFOV PET/CT. While Refs.~\cite{steinberger2024,mercolli2024,mercolli2025} showed that \iod\ outperforms \ga\ and \rb\ in terms of \tge\ count statistics, the current study investigates the performance of \scan\ with respect to oPs lifetime imaging and how it compares to \iod\ using the methodology described in Refs.~\cite{steinberger2024,mercolli2024,mercolli2025}.

%%%%%%%%%%%%%%%%%%%%%%%%%%%%%%%%%%%%%%%%%%%%%%%%%%%%%%%%%%%%%%%%%%%%%%%%%%%%%%%
\section{Method}

\scan\ was produced at the Paul Scherrer Institute (PSI, Switzerland). The radionuclide production and post-irradiation processing at PSI have been established and are being further developed and optimized, as documented in Refs.~\cite{Grundler2020,Braccini2020,vandermeulen2020}.
At Inselspital's Department of Nuclear Medicine (Switzerland) a standard NEMA image quality phantom (Data Spectrum Corp.) without lung insert was filled with a total of $41.7 \,\mbq$ at scan time. The dose calibrator in the Department of Nuclear Medicine (VDC-405/VIK-202, Comecer, The Netherlands) was cross-calibrated with a \scan\ reference activity from PSI. Ref.~\cite{juget2023} describes the calibration of PSI's dose calibrator for \scan. The activity concentration in the six phantom spheres at scan time was $40.68 \,\kbqml$ while the background concentration was $3.90\, \kbqml$. 
The phantom was scanned for $20 \, \mbox{min}$ in the so-called singles mode on a Biograph Vision Quadra (Siemens Healthineers, USA). Singles mode stores all single-crystal interactions into a list mode file. The sorting of \tge\ is performed using the same prototype software as described in Refs.~\cite{steinberger2024,mercolli2024,mercolli2025}. The annihilation photon energy window is 476 to 546 \kev\ with a coincidence time window of 4.2 \ns, while the prompt photon energy window is 720 to 735 \kev, i.e.\ the last two energy bins. No reconstruction algorithm is applied, i.e.\ the spatial localization of the \tge\ is purely based on time-of-flight (TOF) of the 511 \kev\ photons \cite{steinberger2024}. As described in Ref.~\cite{steinberger2024}, Quadra resolves photon energies up to $726\, \kev$. Beyond this energy, all detected photons are collected in a single energy bin. Since the prompt-photon of \scan\ has an energy of $1157.022 \pm 0.015$ \kev, all prompt-photon events are located in the last energy bin. 

The time differences between the annihilation and prompt photons for each \tge\ were binned in order to obtain a PAL spectrum. The time bins are $133 \, \ps$ wide. We followed the Bayesian fitting procedure discussed in Refs.~\cite{steinberger2024,mercolli2024,mercolli2025} in order to obtain the oPs lifetime from the measured PAL spectrum. The PAL spectrum background was determined from time differences smaller than $-2.7 \, \ns$, while the fit was performed for time differences between $-2\,\ns$ and $8.6\,\ns$. The nomenclature and the priors on the fit parameters for this study are defined in Ref.~\cite{mercolli2025}. 

We determined the oPs lifetime for the six spheres $s_{1... 6}$ of the NEMA phantom (nominal diameters: $10$, $13$, $17$, $22$, $28$, $37\,\mbox{mm}$). Furthermore, we binned the spatial distribution of the detected \tge\ into voxels of \imgbin. For each voxel, the oPs lifetime is determined according to the same Bayesian fitting as for the phantom spheres.

%%%%%%%%%%%%%%%%%%%%%%%%%%%%%%%%%%%%%%%%%%%%%%%%%%%%%%%%%%%%%%%%%%%%%%%%%%%%%%%
\section{Results}

\begin{figure}
    \centering
    \includegraphics[width=0.47\linewidth]{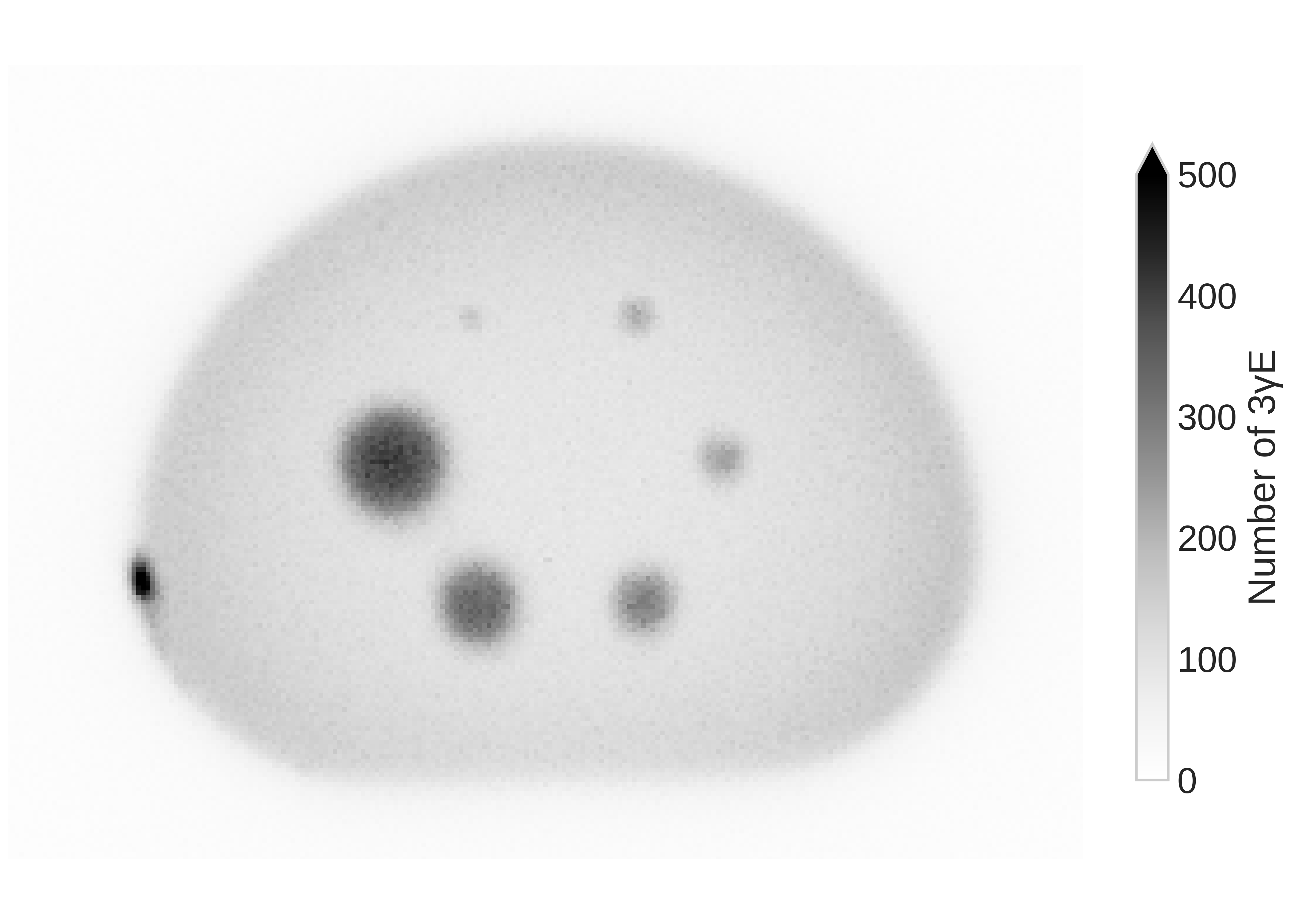}
    \hfill
    \includegraphics[width=0.47\linewidth]{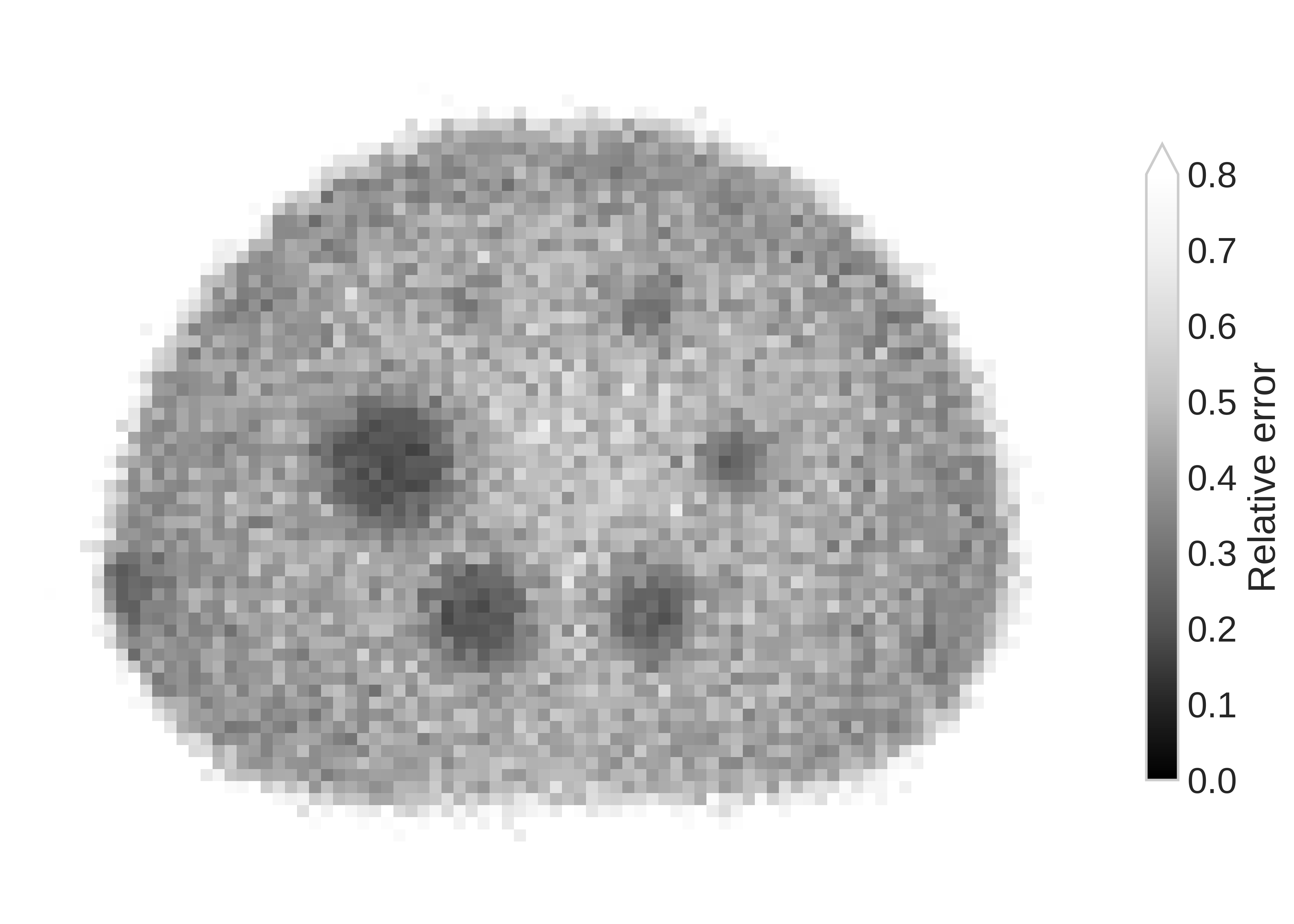}
    \caption{MIP of the histoimage with a voxel size that corresponds to the CT image (left) and the relative error in the background region of the PAL spectrum in a single slice with \imgbin\ voxel size (right). }
    \label{f:histo_img}
\end{figure}

The left panel of Fig.~\ref{f:histo_img} shows the maximum intensity projection (MIP) of the \tge\ histoimage. The binning is chosen according to the CT image, i.e.\ $1.52\times 1.52\times 1.65 \,\mbox{mm}^3$. Even without any reconstruction methodology, i.e.\ using only TOF for the localization of the \tge, the smallest sphere $s_1$ of the NEMA phantom is visible. The absence attenuation correction is clearly visible through the darkening on the border of the phantom. Some \scan\ activity stuck to the left wall of the phantom. 

On the right of Fig.~\ref{f:histo_img} the relative error in the background region of the PAL spectrum, i.e.\ for time differences that are smaller than $-2.7\,\ns$, is shown. The error inside the spheres decreases as there is a higher activity concentration. Due to the decreasing number of \tge\ towards the center of the phantom, the error increases towards the center of the phantom (there is no attenuation correction).  

\begin{figure}
    \centering
    \includegraphics[width=0.85\linewidth]{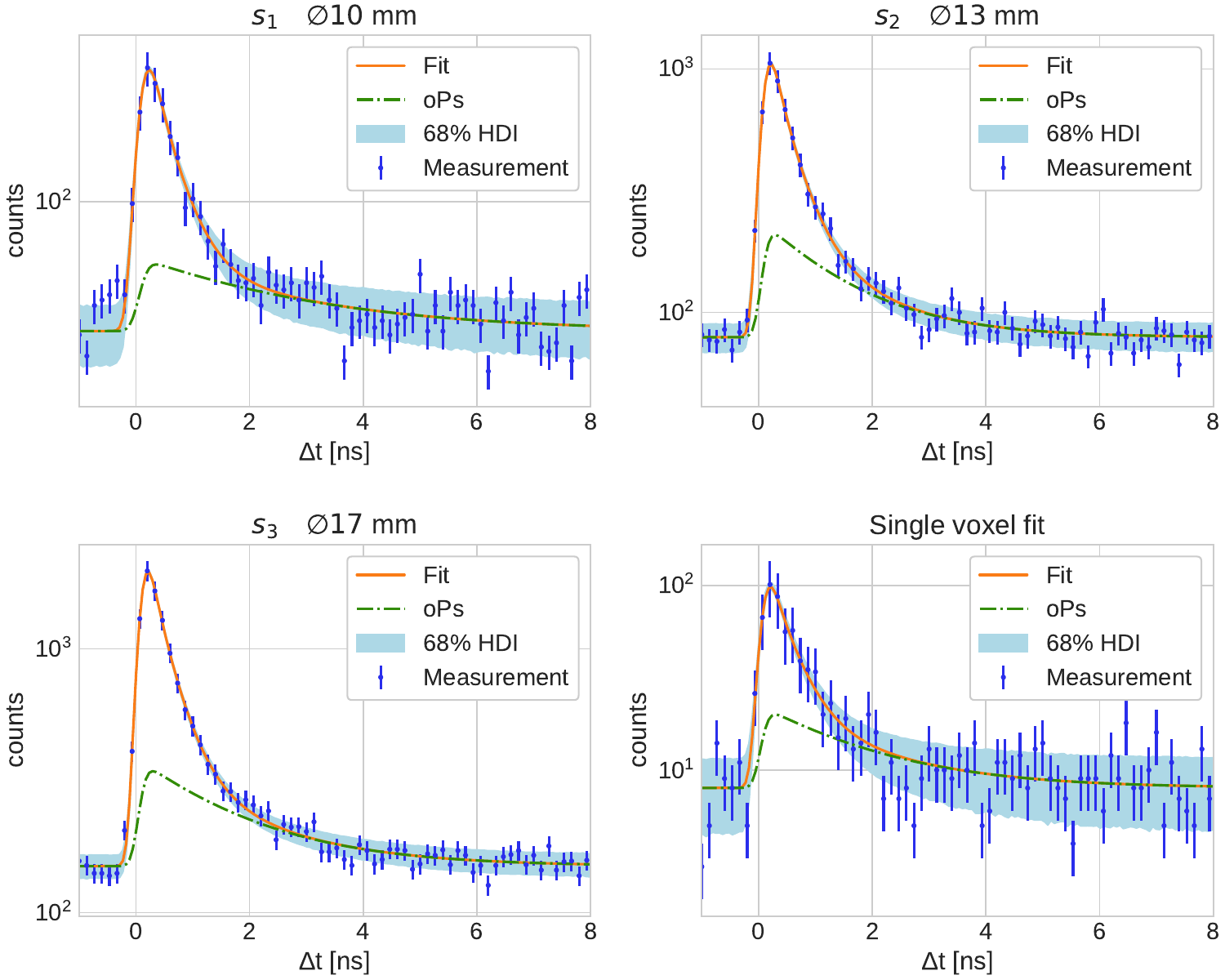}
    \caption{PAL spectrum of all \tge\ with the fit prediction in the three smallest spheres of the NEMA phantom and of a single \imgbin\ voxel in the center of $s_6$.}
    \label{f:fit_spheres}
\end{figure}

Fig.~\ref{f:fit_spheres} shows the measured PAL spectrum with the fit prediction for the three smallest spheres and a single voxel in the center of the largest sphere $s_6$. The error bars plotted on the measurement points are the relative error in the background region of the PAL spectrum, i.e.\ the relative standard deviation of all time differences $<-2.7 \, \ns$. %We checked that the statistical error in the PAL spectrum does not follow a Poisson distribution and we therefore take the standard deviation of the time differences $<2.7 \, \ns$. 
The $68\%$ highest density interval (HDI) plotted in Fig.~\ref{f:fit_spheres} represents prediction uncertainty of the fit. The fit results corresponding to the PAL spectrum in Fig.~\ref{f:fit_spheres} are reported in Tab.~\ref{t:fit_spheres} together with the fit results of the larger phantom spheres. The posterior distribution of $\tops$ is Gaussian, hence we report the error on $\tops$ as a standard deviation in Tab.~\ref{t:fit_spheres}. This does not apply to $BR_{1,2,3}$, which is why their error is quoted as a $68\%$ HDI. 

\begin{table}[]
    \centering
    \caption{Fit results for the six phantom spheres and a single \imgbin\ voxel in the center of $s_6$. } \label{t:fit_spheres}
    \pgfplotstabletypeset[col sep=semicolon,
        columns/Fit/.style={column type=l,column name=Fit, string type},
        columns/τoPs/.style={column type=l,column name=$\tops \, [\ns]$, string type},
        columns/HDIτ/.style={column type=l,column name=$\mbox{HDI}_{\tops} \, [\ns]$, string type},
        columns/BR1/.style={column type=c,column name=$BR_1$, string type},
        columns/HDIBR1/.style={column type=c,column name=$\mbox{HDI}_{BR_1}$, string type},
        columns/BR2/.style={column type=c,column name=$BR_2$, string type},
        columns/HDIBR2/.style={column type=c,column name=$\mbox{HDI}_{BR_2}$, string type},
        columns/BR3/.style={column type=c,column name=$BR_3$, string type},
        columns/HDIBR3/.style={column type=c,column name=$\mbox{HDI}_{BR_3}$, string type},
        every head row/.style={before row=\toprule,after row=\midrule},
        every last row/.style={after row=\bottomrule},
        every even row/.style={before row=\rowcolor{lightgray}},
    ]{Fit;τoPs;BR1;HDIBR1;BR2;HDIBR2;BR3;HDIBR3
$s_1 \; \varnothing 10 \,\mbox{mm}$;$2.65\pm0.50$;0.072;[0.0, 0.091];0.659;[0.608 0.736];0.269;[0.242 0.301]
$s_2 \; \varnothing 13 \,\mbox{mm}$ ;$1.39\pm0.20$;0.077;[0.049, 0.106];0.623;[0.573, 0.679];0.30;[0.267 0.324]
$s_3 \; \varnothing 17 \,\mbox{mm}$;$1.76\pm0.18$;0.062;[0.041, 0.083];0.651;[0.62, 0.687];0.287;[0.27  0.301]
$s_4 \; \varnothing 22 \,\mbox{mm}$;$1.86\pm0.09$;0.057;[0.047 0.067];0.655;[0.639 0.671];0.288;[0.281 0.296]
$s_5 \; \varnothing 28 \,\mbox{mm}$ ;$1.73\pm0.1$;0.091;[0.08  0.103];0.603;[0.585 0.622];0.306;[0.296 0.314]
$s_6 \; \varnothing 37 \,\mbox{mm}$;$1.78\pm0.08$;0.066;[0.057 0.076];0.642;[0.627 0.657];0.292;[0.285 0.299]
Voxel;$1.79\pm0.57$;0.051;[0.0, 0.063];0.609;[0.553, 0.717];0.34;[0.266, 0.386]
}
\end{table}

\begin{figure}
    \centering
    \includegraphics[width=0.47\linewidth]{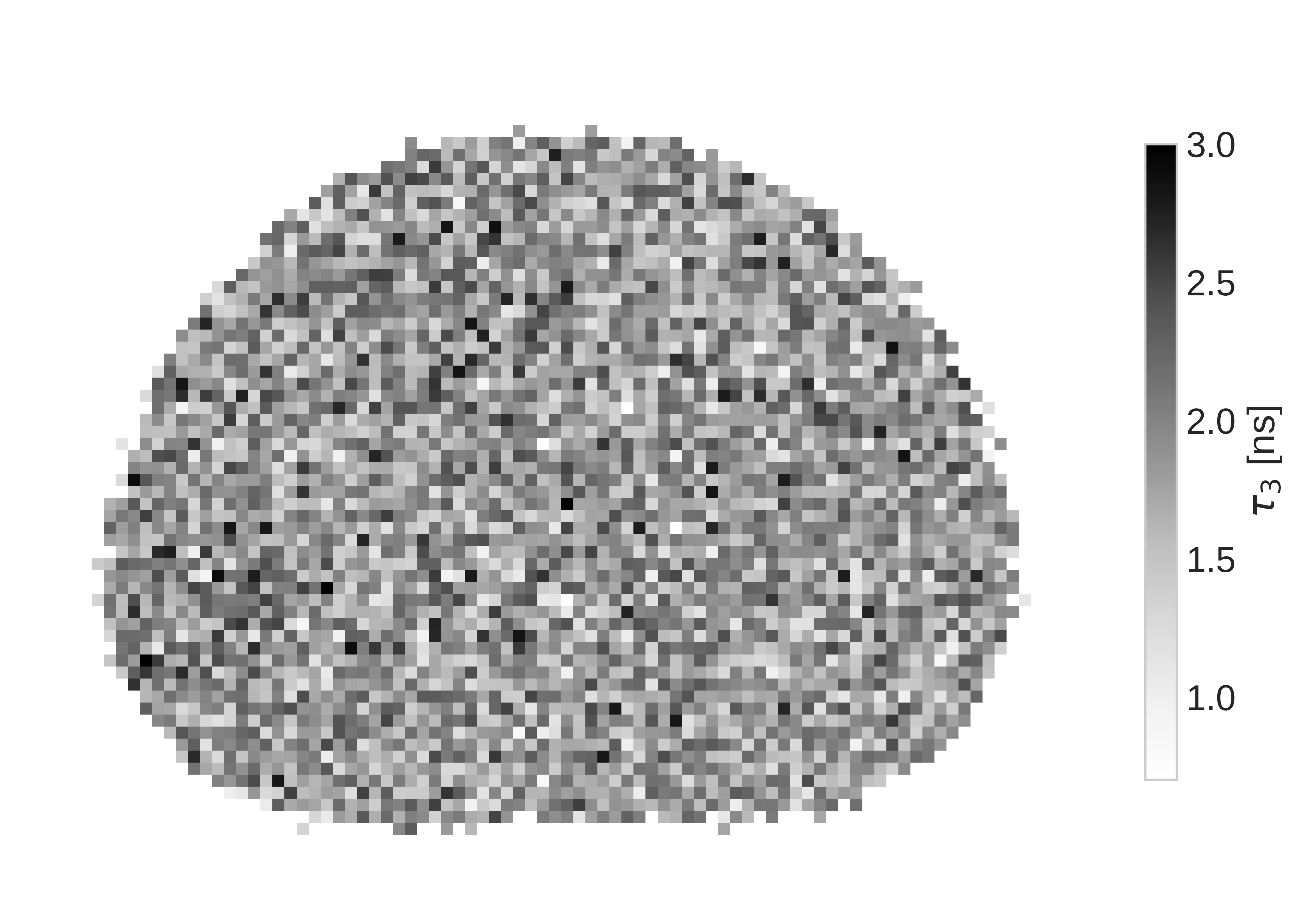}
    \hfill 
    \includegraphics[width=0.47\linewidth]{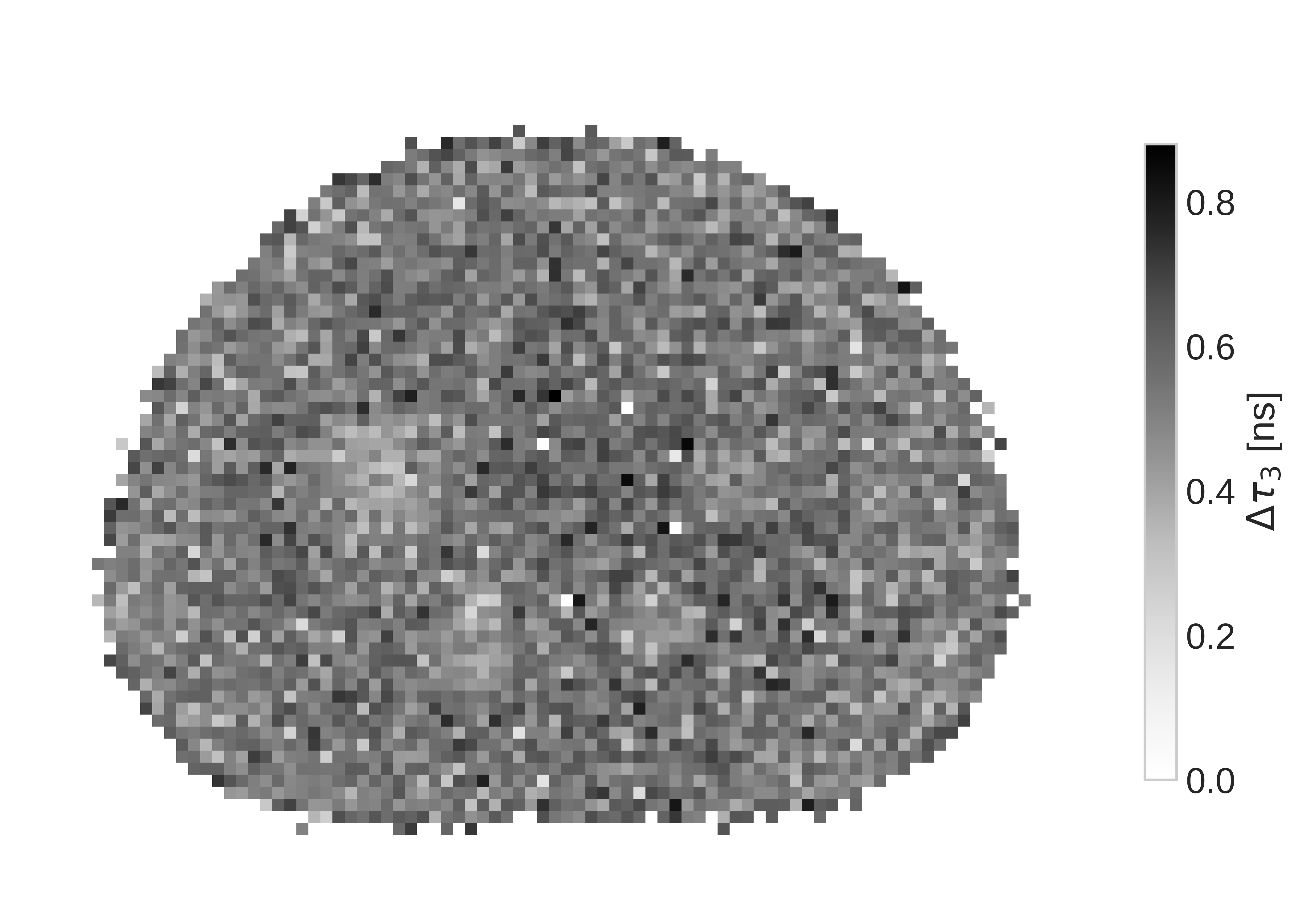}
    \caption{Slice of the oPs lifetime image (left) and $\tops$ error (right) with \imgbin\ voxels. }
    \label{f:ops_imaging}
\end{figure}

In Fig. \ref{f:ops_imaging} a slice of the full oPs lifetime image, together with the fit error on $\tops$ with a \imgbin\ binning, is presented. While the oPs lifetime image is not particularly interesting - after all, the phantom is filled with water - the marginalized uncertainty on $\tops$ clearly increases in the central region of the phantom. Note that only for the four largest spheres, the error decreases visibly.

%%%%%%%%%%%%%%%%%%%%%%%%%%%%%%%%%%%%%%%%%%%%%%%%%%%%%%%%%%%%%%%%%%%%%%%%%%%%%%%
\section{Discussion}

From the discussion in Ref.~\cite{steinberger2024}, it is clear that the key question is whether the high prompt photon BR of \scan\ can overcome the Quadra's inability to resolve \scan's photopeak. Detector hits above $726 \,\kev $ are collected in a single integrating bin. One should, therefore, expect that more random coincidences are selected due to the high prompt photon energy of \scan. The right panel of Fig.~\ref{f:fit_spheres} already hints towards a high random \tge\ rate: even inside the spheres, the relative error in the background region of the PAL spectrum exceeds $20\%$. For a comparison, Ref.~\cite{mercolli2025} only considered those voxels with less than $20\%$ background error for oPs lifetime imaging.

The large number of random \tge\ is reflected in the statistical uncertainty of $\tops$ reported in Tab.~\ref{t:fit_spheres}. All values for $\tops$ in the phantom are consistent with the literature value of $1.839 \pm 0.015 \,\ns$ for water from Ref.~\cite{kotera2005} and with the results from Ref.~\cite{mercolli2025} within their statistical uncertainty (note also the reference values in Ref.~\cite{Stepanov2020}). However, the marginalized uncertainties reported in Tab.~\ref{t:fit_spheres} are rather large: only starting from $s_3$ the relative error starts dropping below $10\%$ (and reaches even $31.9 \%$ in a single voxel). This is likely more than the precision required to sense different oxygenation levels in lesions, as discussed in Ref.~\cite{shibuya2020}). 

$\tops$'s uncertainty is seen in Fig.~\ref{f:ops_imaging} as well. The variation on $\tops$ across the whole phantom is quite large, given that the expected oPs lifetime should be the same across the whole phantom. In the right panel of Fig.~\ref{f:ops_imaging}, only very few voxels have an error below $0.3 \, \ns$. The mean uncertainty on $\tops$ across the slice shown in Fig.~\ref{f:ops_imaging} is $0.53 \,\ns$. Only the four largest spheres of the phantom have a visibly smaller uncertainty compared to the phantom background. 

The fit of the oPs lifetime critically depends on the time differences after the peak in the PAL spectrum, i.e.\ on values close to the random \tge\ background. A useful quantity to characterize the \tge\ count statistics is therefore the peak signal-to-background ratio (pSBR) in a PAL spectrum. In the measurements with \iod, Ref.~\cite{mercolli2025} reported a pSBR of about 55.5 for a \imgbin\ voxel in the water tube with an activity concentration of $252 \, \kbqml$ and a scan time of $15 \, \mbox{min}$. For the PAL spectrum in the \imgbin\ voxel in Fig.~\ref{f:fit_spheres}, however, the pSBR is only about 12.6. Despite the activity concentration being higher in the \iod\ measurements of Ref.~\cite{mercolli2025}, the scan duration is $5\,\mbox{min}$ shorter.
The error on $\tops$ in a single voxel (last row in Tab.~\ref{t:fit_spheres}) is about four times larger than the error reported in Ref.~\cite{mercolli2025} for the same voxel size. %Assuming only a scaling with the square root of the activity concentration, i.e.\ without considering the scan duration, prompt photon BR, or any other effect, the error on $\tops$ from \scan\ should be about 2.5 times larger than from \iod. 
A similar picture arises when looking at volumes of similar size, e.g.\ the sphere $s_4$ has a volume of $5.57 \, \mbox{ml}$ and is comparable with the volume of the tubes in Ref.~\cite{mercolli2025}. The relative error on $\tops$, however, is $4.8 \,\%$ while Ref.~\cite{mercolli2025} reports a $1.1,\%$ error for a $5\,\mbox{ml}$ tube with water. This comparison is even more striking, when considering the prompt-photon BR per positron, which is almost 8 times higher for \scan\ than for \iod. 
We conclude that, with the given methodology, resolving the photopeak is key for a low random \tge\ rate. \scan's high prompt photon BR cannot overcome Quadra's limited detection capabilities for high-energy photons. 
It should be emphasized that this conclusion applies to the given methodology. Different detection methods~\cite{moskal2024brain} or event selection procedures and/or random \tge\ estimations as e.g.\ in Ref.~\cite{Huang2024} may reduce the uncertainties on $\tops$ in the case of high-energy prompt photons. We leave such an investigation for future studies. 

In contrast to \scan, \isotope[43]{Sc}'s prompt photon is within Quadra's energy range (see Tab.~\ref{t:isotopes}) and therefore, the afore mentioned discussion of the high-energy prompt-photons does not apply. However, the prompt-photon BR per positron is in the same order of magnitude as \iod\ and \rb, i.e.\ much lower than for \scan. 

% So far we did not consider the positron energy of the radionuclides in Tab.~\ref{t:isotopes}. The positron range prior to positronium formation or direct annihilation is important for image quality. \iod\ is clearly limited (see e.g.\ ) . The discussion of positron range effects, however, is only meaning full in the context of proper image reconstruction for oPs lifetime imaging (see Refs.~\cite{qi2022,Huang2024}  a discussion, however, is 

%%%%%%%%%%%%%%%%%%%%%%%%%%%%%%%%%%%%%%%%%%%%%%%%%%%%%%%%%%%%%%%%%%%%%%%%%%%%%%%
\section{Conclusions}

Given Quadra's limited energy resolution and the current methodology for selecting \tge, it does not seem that \scan\ is able to outperform \iod\ in terms of count statistics for oPs lifetime imaging, despite its favorable physical properties and clinical prospects.%, \scan\ is not able to outperform \iod\ for oPs lifetime imaging despite its favorable physical properties and clinical prospects. 

%%%%%%%%%%%%%%%%%%%%%%%%%%%%%%%%%%%%%%%%%%%%%%%%%%%%%%%%%%%%%%%%%%%%%%%%%%%%%%%
%%%%%%%%%%%%%%%%%%%%%%%%%%%%%%%%%%%%%%%%%%%%%%%%%%%%%%%%%%%%%%%%%%%%%%%%%%%%%%%

\section*{Conflict of Interest Statement}

WMS and MC are full-time employees of Siemens Medical Solutions USA, Inc. HS is a part-time employee of Siemens Healthineers International AG. 
PM is an inventor on a patent related to this work. Patent nos.: (Poland) PL 227658, (Europe) EP 3039453, and (United States) US 9,851,456], filed (Poland) 30 August 2013, (Europe) 29 August 2014, and (United States) 29 August 2014; published (Poland) 23 January 2018, (Europe) 29 April 2020, and (United States) 26 December 2017. AR has received research support and speaker honoraria from Siemens. KS received research grants from Novartis and Siemens and conference sponsorships from United Imaging, Siemens, and Subtle Medical not related to the submitted work. All other authors have no conflict of interests to report.

\section*{Author Contributions}

LM conceptualized the study, carried out the data collection, performed the data analysis and wrote the manuscript. WMS performed part of the data analysis. PVG, AM and NPvdM produced the \isotope[44]{Sc} and wrote parts of the manuscript. All other authors helped in some capacity for coordinating, planning/executing the experiments or understanding the results. All authors read and approved the manuscript.

\section*{Funding}
This research is partially supported by the grant no. 216944 under the Weave/Lead Agency program of the Swiss National Science Foundation and the National Science Centre of Poland through grant OPUS24+LAP No. 2022/47/I/NZ7/03112 and 2021/43/B/ST2/02150. The dangerous good transportation was financed by the Research Fund of the Swiss Society of Radiobiology and Medical Physics.

% \section*{Acknowledgments}

% % \section*{Supplemental Data}
% %  \href{http://home.frontiersin.org/about/author-guidelines#SupplementaryMaterial}{Supplementary Material} should be uploaded separately on submission, if there are Supplementary Figures, please include the caption in the same file as the figure. LaTeX Supplementary Material templates can be found in the Frontiers LaTeX folder.

\section*{Data Availability Statement}

The evaluated data used in this study is available upon reasonable request from the corresponding author.

%%%%%%%%%%%%%%%%%%%%%%%%%%%%%%%%%%%%%%%%%%%%%%%%%%%%%%%%%%%%%%%%%%%%%%%%%%
\newpage
\printbibliography

\end{document}